\documentclass[twocolumn,preprintnumbers,amsmath,amssymb]{revtex4}

\usepackage{graphicx}
\usepackage{dcolumn}
\usepackage{bm}

\begin{document}

\preprint{NStaley}

\title{Exploring geometrical constraint effects of Cooper pairs}

\author{Neal E. Staley}
\author{Ying 
Liu}
\email{liu@phys.psu.edu}
\affiliation{%
Department of Physics, 
The Pennsylvania State University, University Park, PA 16802, 
U.S.A.
}%
\date{\today}

\begin{abstract}
We report low-temperature electrical transport measurements on filled Al squares prepared by e-beam lithography featuring a sample size ranging from 130 nm to 530 nm. The values of the superconducting coherence length, $\xi$(0), were found to range from $\sim$100-210 nm. We found that phase diagrams in the magnetic field and temperature space feature quantized steps in samples with a size comparable with $\xi$(0), suggesting the emergence of quantized states in these smallest superconducting structures. These quantized states, which are not anticipated in Ginzburg-Landau theory, may be associated with the quantum size effects of Cooper pairs.
\end{abstract}

\maketitle

Shortly after the publication of the Bardeen, Cooper, Shrieffer (BCS) theory of superconductivity, Anderson pointed out that superconductivity is fully suppressed when a superconducting grain becomes so small that it's electron energy level spacing approaches the bulk superconducting energy gap \cite{Anderson59}, a phenomenon known as the quantum size effect of superconductivity. Experimentally, the superconducting energy gap of Al grains were found to indeed fall rapidly when the diameter of the grain was reduced to a critical size of L$_c$  (material dependent, $\sim$6-10 nm for Al)\cite{TinkhamReview}. The value of L$_c$ was found to also depend strongly on whether the number of electrons in the grain is odd or even\cite{BlackTinkhamNmAl}. The interesting question is how Cooper pairs will respond to geometrical constraints in a grain when its size is larger than L$_c$ but smaller than or comparable with the size of a Cooper pair, which in BCS theory is the superconducting coherence length, $\xi_0$. The corresponding length in Ginzburg-Landau (G-L) theory is $\xi$(0), which characterizes the size of the normal core of an Abrikosov vortex and is the same order of magnitude as $\xi_0$. In the presence of disorder, the Cooper pair size is reduced  - $\xi_0$ becomes $(\xi_0l)^{1/2}$ and is related to $\xi$(0) by $\xi(0) = 0.74(\xi_0l)^{1/2}$, where $l$ is the mean free path. Geometrical constraint effects of Cooper pairs my be expected in these ultrasmall superconductors.  

Over the past twenty years, much work on small superconductors was focused on samples with a size several times of $\xi$(0), and thus much larger than the size of a Cooper pair. In singly connected samples such as filled squares or disks, in the presence of an external magnetic field, their behavior is determined by few vortex physics\cite{BuissonPhysLetA, MoshcalkovNature95, GeimNature97, ChibotaruNatureAntivortex}. More recently, superconductors with a size smaller than or comparable with $\xi$(0) were studied primarily by a scanning tunneling spectroscopy probe \cite{NishioSTMPb, WangPseudogapPbSTM, BoseShellEffectsSnSTM}. It was found that in ultra small singly connected samples no vortices can exist below a critical size.  Also, an unexpectedly large pseudogap related to modified electron-phonon interaction exists, as well as a quantum size effect enhanced superconducting gap. In addition, doubly connected samples have been studied, showing that the fundamental fluxoid quantization, which is responsible for the well-known Little-Parks oscillations\cite{LittleParks}, leads to emergence of a ñdestructive regimeî for ultrasmall superconducting loops with a diameter d less than $\xi$(0), as predicted by de Gennes \cite{deGennesLoop}. Here the superconducting transition temperature of a loop with its circumference less than $\pi \xi$(0) is suppressed to zero near half-integer flux quanta\cite{deGennesLoop}, as confirmed experimentally\cite{LiuScience01}. In all these cases, however, the issue of the geometrical constraint effects of Cooper pairs was not addressed.

In the present work we explore possible quantum size effects of Cooper pairs in a singly connected superconducting square with its size similar to $\xi$(0). Within G-L theory, a phenomenological framework describing superconductivity, no vortex and, furthermore, no spatial variation in the superconducting energy gap is allowed in isolated ultrasmall superconducting samples of this size under any applied magnetic fields. Therefore no quantized states in the superconductor are expected in G-L theory. Consequently, the phase diagram in magnetic field $vs.$ temperature (H-T) should be featureless\cite{SchweigertPeeters98}. On the other hand, if we view the singly connected sample as being composed of a series of doubly connected loops, a suitable magnetic field should place some inner loops in the destructive regime but leave the outer loops superconducting, leading to spatial variation of the superconducting energy gap and possible quantized states.

The question of the existence of quantized states in ultrasmall superconductors can be addressed experimentally. Recent advances in nanofabrication techniques have enabled the fabrication of devices with a size needed to address this issue.  Because the longest $\xi$(0) in thin films tends to be on order 100nm, we fabricated our devices using ebeam lithography with a single layer PMMA resist to achieve smaller feature sizes.  All writes were performed using a Leica EBPG5-HR ebeam writer.   Because we were interested in obtaining long coherence length films, we thermally deposited 20nm of 99.9999\% (6N) purity aluminum, and defined the devices by a liftoff procedure.  Devices were fabricated in a quasi-four point geometry (see Fig. 1a for a schematic) where the transport included a portion of thin Al leads as well as the confined superconducting Al square.  It should be noted that we rely on the difference in the widths of the leads and square, rather than normal leads, to provide the confining geometry. After fabrication we imaged each device using a scanning electron microscope (SEM) to determine the geometry.  We constructed critical field verses temperature phase diagrams of each sample from resistivity measurements taken via a dc technique in an RF filtered dilution refrigerator.

\begin{figure}
\includegraphics[ scale 
=1.0]{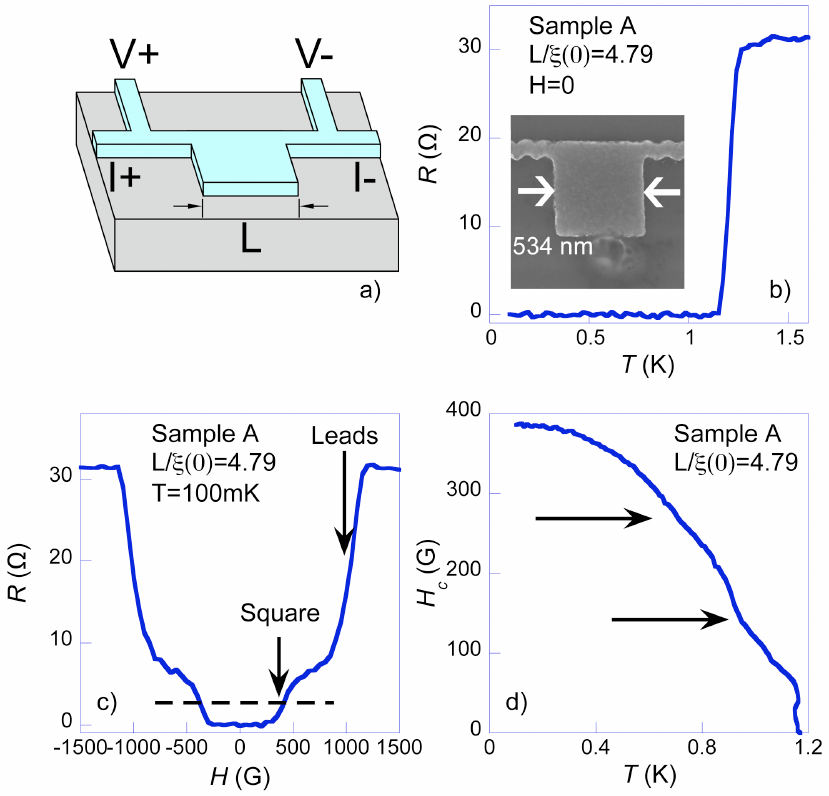}
\caption{(Color Online) a) Schematic of Al devices used in this study.  The width and the length of the current and voltage leads to the square vary between devices and connect to large Al contact pads;  b) Resistance ($R$) $vs$. temperature ($T$), $R$($T$), at zero magnetic field ($H$ = 0) showing a superconducting transition temperature ($T_c$) of 1.20 K. Inset: Scanning electron microscope (SEM) image of this $\sim$500 nm square sample.  Note that the wavy contact lines were due to an old filament installed on the Leica EBPG5-HR e-beam writer and the high doses used for writing this device. The organic contaminants shown at the bottom center of the image should not affect the device; c) Resistance $vs$. magnetic field applied perpendicular to the substrate surface, $R$($H$).  The high and low critical fields correspond to those of the thin leads and the square, respectively; d) Phase diagram constructed from $R$($H$) and $R$($T$) data.  Arrows indicate the magnetic fields predicted for the transitions of few vortex states in a circular disk of the same area as the square (see text).}
\end{figure}

The phase diagram in $H$-$T$ space can be obtained by searching for the appropriate magnetic field/temperature values for which the sample resistance was kept at a fixed value, while the temperature and the magnetic field were swept. Because the sample has multiple transitions, due to the differing sizes of the leads and square, measurements of resistance $vs$. magnetic field at selected temperatures were used to find the resistance value corresponding to the superconducting transition of the square.  This allowed us to construct the $H$-$T$ phase diagram for each of the superconducting squares from our transport measurements.   Note the convenience of this technique in that it allows the determination of the phase diagram even when one or two leads are lost by taking into account the fixed resistance offset of the leads. In Figure 1, we show results for Sample A that consists of a square roughly 534 nm across and two 94 nm wide leads. In addition, two voltage leads are located 600 nm away from the square (not shown). The sample has a superconducting transition temperature (T$_c$) of 1.20 K in zero field, close to the bulk T$_c$ for Al.  To estimate the values of $\xi$(0) and the penetration depth $\lambda$(0), we measured both the perpendicular critical field H$_{c\perp}$, using a bulk film from the same run, and the parallel critical field H$_{c\parallel}$ using the thin leads connecting to the square, estimating $\lambda$(0) using\cite{DeGennes} 
\begin{equation}
H_{c\parallel}(T)= 2\sqrt{6} \frac{H_{c}(T)\lambda_{eff}(T)}{d}
\end{equation}
and coherence length from the equation
\begin{equation}
\xi(T)=\frac{\Phi_0}{2\sqrt{2}\pi H_{c}(T)\lambda_{eff}(T)}
\end{equation}
where $\Phi_0$ = $\frac{hc}{2e}$ is the flux quantum, $d$ is the apparent thickness of the parallel portion and H$_c$ is the thermodynamic critical field.  Combining these two equations we can estimate $\xi$(T) by simply measuring H$_{c\parallel}$ valid for both type I and type II superconductors
\begin{equation}
\xi(T)=\frac{\sqrt{3}\Phi_0}{\pi d Hc_\parallel (T)}
\end{equation}
If our films are type II superconductors, H$_{c\perp}$= H$_{c2}\neq$ H$_c$ implying that we cannot accurately estimate $\lambda$ however we can estimate $\xi$(T) from the measured H$_{c2}$ by
\begin{equation}
H_{c2}=\frac{\phi_0}{2\pi \xi^2(T)}
\end{equation}
We can calculate both $\lambda(0)$ and $\xi(0)$ by taking into account the temperature dependance $\xi(T)^2=\xi(0)^2\frac{1}{1-\frac{T}{T_c}}$ and $\lambda(T)^2=\lambda(0)^2\frac{1}{1-\frac{T}{T_c}}$.  From the experimental value of H$_{c\perp}$ = 175 G and H$_{c\parallel}$ = 1166 G at 100 mK, we obtained a $\xi(0)$ of 112 nm from the H$_{c\parallel}$ and estimate $\lambda$(0) of 128 nm, noting that this implies the sample is type II and the value of $\lambda$(0) from this estimate is not valid.  Since the film is type II we also estimated $\xi$(0) from H$_{c\perp}$ to be 131nm.  It should be pointed out that in the above estimate was taken for a film evaporated at the same time as the squares, so should be similar to $\xi$(0) for the samples, while the H$_{c\parallel}$ estimation for $\xi$(0) was taken for the thin leads connected to the square which should better represent the properties of the film at the square.  Nevertheless, because the values of $\xi$(0) estimated here are used only to characterize the samples, and are not used in any calculation, the rough estimates do not have any serious consequence on our analysis of the results.  For the sample shown in Fig. 1, $L/\xi(0)=4.79$, where $L$ is the size of the square.  In Fig. 1d, we show the $H$-$T$ phase diagram constructed from $R$($H$) measurements taken at various temperatures.  

Several interesting features are seen in the phase boundary shown in Fig. 1d indicating possible transitions between different states. In addition to a prominent ``bump" near 70 G, two relatively subtle ``kinks" are also seen near 145 and 270 G, respectively. In order to understand the physical origin of these features, it is useful to examine the expectations from G-L theory for the modulation of T$_c$ as a function of applied magnetic field. For disks larger than a critical size the solution to the G-L equation features eigenvalues of angular momentum corresponding to the mesoscopic disk admitting vortices. While the analytical solution for squares is not available, numerically derived phase diagrams for vortex decoration exist in the literature\cite{MerteljVortexPhaseDiagram}, but don't address the low field giant vortex transitions.   However the single vortex and giant vortex solutions are expected to occur at similar  fields as in a disk of equal area.  We estimated the transition fields for n-vortex states for a disk with area equal to that of the measured square based on the analytic solutions obtained previously\cite{BenoistZwerger96}.  For disks of this area, the expected $n$=0 to $n$=1 transition should occur at 145 G and the $n$=1 to $n$=2 transition should occur at 260 G. Therefore the two ``kinks" found in the phase diagram near these magnetic fields must be associated with the few-vortex states of the square, similar to previous measurements\cite{ChibotaruNatureAntivortex}.  On the other hand, no feature is expected near 70 G within the G-L theory. The physical origin of this ``bump" is not understood.  

The observation of few-vortex states in moderately confined Al squares is reasonable, because the sample size is larger than the size of a vortex.  However as the sample size is reduced to be comparable with the size of the normal core of a vortex, roughly twice $\xi$(T), this vortex state should cease.  Indeed, the solution of the G-L equation for a superconducting disk indicates that when the radius falls below a critical value of 1.319$\xi$ \cite{SchweigertPeeters98}, the disk remains vortex free for all magnetic fields below the intrinsic critical field. It is important to note that this is the regime where G-L theory, which can describe physics at a length scale larger then $\xi(0)$ (all physical quantities in the G-L theory are averaged over this length), starts to fail. Therefore how the superconducting order parameter responds to an applied magnetic field can not be treated self-consistently by G-L theory in this regime. Experimentally, to approach this regime either the sample size needs to be reduced, or the coherence length increased, or both. We show in Fig. 2, a 546 nm square prepared with $\xi$(0)= 209 nm, corresponding to L/$\xi(0)$=2.61, which has no features identifiable with few-vortex states in it's phase diagram (Fig. 2a), indicating that it was in the $n$=0 state for the whole temperature range probed.  Similarly for a 250 nm square with $\xi$(0)=102 nm, corresponding to L/$\xi(0)$=2.38, no vortex state was observed.  These observations suggest that the Al square is too small to support a single vortex, as expected.

\begin{figure}
\includegraphics[ scale 
=1.0]{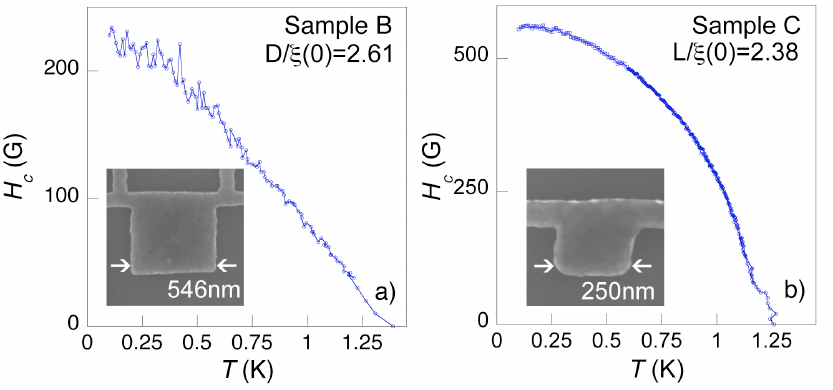}
\caption{(Color Online) Critical field versus temperature for two squares with their sizes and estimated $\xi(0)$ indicated.  Insets show SEM images of each sample.}
\end{figure}

We reduced the relative size of the Al squares to $\xi(0)$ to further understand the effect of confinement on the Cooper pairs.  In Fig. 3, the phase diagrams of two Al squares featuring sizes of 240 nm and 138 nm, a zero field T$_c$ of 1.23 K and 1.28 K and a $\xi$(0) of 112 nm and 117 nm, respectively, are shown. Given that L/$\xi$(0)=2.13 and L/$\xi$(0)=1.18, a smooth phase boundary similar to those shown in Fig. 2 is expected from G-L theory. Surprisingly, sharp steps are seen in the phase diagrams of these two samples.  In Fig. 3a, a sharp rise in critical field H$_c$ is seen as the temperature is lowered to below 0.9 K and a less distinct rise at 1.23 K (near T$_c$) as well.  For the sample shown Fig. 3b featuring even stronger confinement than that shown in Fig. 3a, two pronounced, very sharp steps are seen, occurring at $T$ = 1 K and 0.6 K.  The systematic behavior found in these two samples suggests strongly that these step features are intrinsic features of the Al devices. 

\begin{figure}
\includegraphics[ scale =1.0]{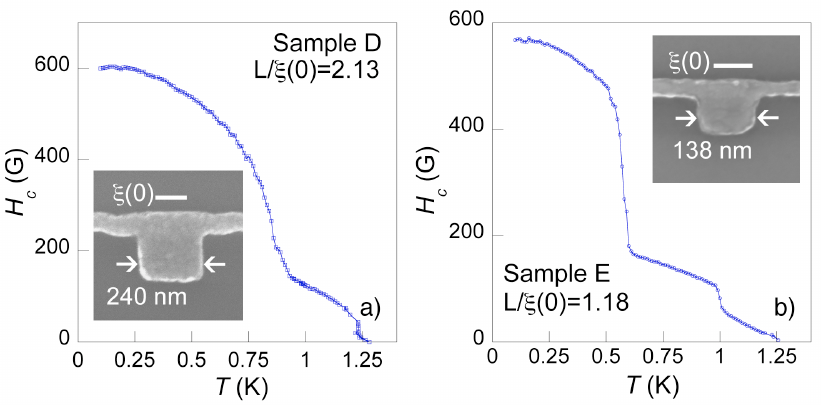}
\caption{(Color Online) Critical field versus temperature for a two ultrasmall Al squares, a nominally 240 nm square with $\xi$(0) = 110 nm (a) and a nominally 140 nm square with a $\xi$(0) = 116 nm. The phase diagram for Sample D was obtained by three-point measurements, and Sample E by two-point measurements. Insets show SEM images of the samples with the relevant length scales.}
\end{figure}

The presence of measurement leads in our sample, however, makes even the smallest filled square we studied an open system. As a result, we treat the square as well as the leads as a whole system in our attempt to obtain insight into the origin of the steps seen in the phase diagram.  For Sample E, $\xi$(0) of the superconducting leads is 116 nm and the square size is 138 nm, making it reasonable to treat both the square and the leads as a quasi one-dimentional (1D) wire featuring two widths and branches associated with voltage leads. In the zeroth order approximation, we can ignore the complications of these leads, and make use of Eq. 3 to relate measured H$_{c\parallel}$ of the square to $\xi$(T) within the G-L theory even though the exact value of $\xi(0)$ so deduced may not be accurate. Noting that $\xi(T) \sim(1-T/T_c)^{-1/2}$ we can plot $\xi(T)(1-T/T_c)^{1/2}$ $vs.$ $T$, as shown in Fig. 4 for sample E, showing two strikingly flat steps. Note that while the functional form $\xi$(0)=$\xi(T)(1-T/T_c)^{1/2}$ is only strictly valid near T$_c$ the flat steps far from T$_c$ still signify a change in $\xi$(0).  The implications of Fig. 4 is that the quantized states seen in Fig. 3 may be associated with different values of $\xi$(0).  Essentially, as the temperature is lowered to below T$_c$, the sample starts off with a very long intrinsic coherence length that in fact diverges at T$_c$, which reduces to a $\xi(0)\sim$ 800 nm near 1 K, then jumps to a stable state with $\xi(0)$ $\sim$ 408 nm and again at 0.55 K to another state with $\xi(0) \sim$ 150 nm. Interestingly, the same plot of the data for Sample D was found to result in very similar results, even though the size of the square is much larger than $\xi(0)$, making the above argument not applicable.

\begin{figure}
\includegraphics[ scale =1.0]{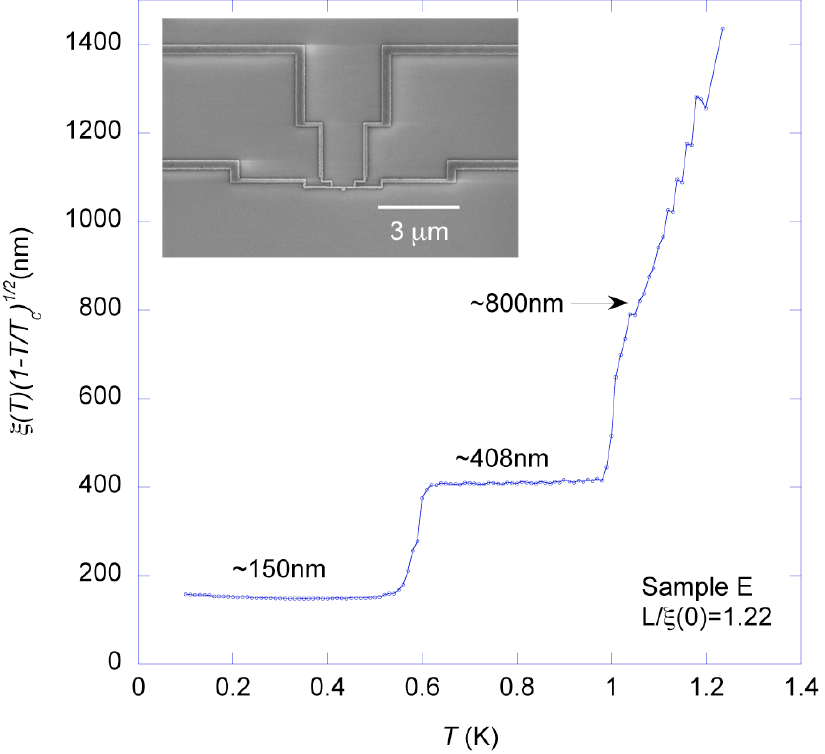}
\caption{(Color Online) $\xi(T)(1-T/T_c)^{1/2}$ plotted as a function of temperature for Sample E with T$_c=1.28$K. Inset shows large scale SEM image of this device, note the different width leads.  Apparent streaking in the image due to charging of the substrate.  
}
\end{figure}

This suggests that the observed steps in the phase diagram might be associated with geometrical constraint effects of superconductivity in ultrasmall Al squares. Consider now a quasi 1D sample that has exactly the same geometry as the experimental sample except that the square width is reduced to be the same as the measurement leads. Obviously no geometrical confinement effects are expected. In this limit, in zero field, the superconducting order parameter will remain constant throughout the sample, with the exception that it may be enhanced near the nodes of voltage leads\cite{FinkSideBranch}. With the addition of a moderate magnetic field the order parameter should remain constant except over a length scale of $\xi$(T) near the bulk leads where it will be smoothly varying to accommodate them being driven normal. As the temperature increases, $\xi$(T) increases and diverges at T$_c$ but no abrupt changes are expected. Now if we start to enlarge a section of the wire gradually and eventually recover the original sample geometry featuring a filled square, the order parameter is expected to evolve correspondingly, especially once the enlarged section becomes thick enough to provide a ``trap'' for Cooper pairs. The square and the rest of quasi 1D system should still feature a well-defined order parameter smoothly varying near the joint between the quasi 1D part of the wire and the square. Since the square can not host a vortex, the observed quantization could be due to the quantum-size effects of Copper pairs. With $\xi$(0) being essentially the size of the Cooper pair, quantized values of $\xi$(0) could be associated with quantization of the Cooper pair size. Our experimental results may suggest that the trapped superconducting condensate adopts different patterns of the superconducting order parameter characterized by different values of $\xi$(0) as the temperature is raised.

This analysis, that Cooper pairs can exhibit different patterns having different values of $\xi$(0), assumes that our geometry can ``trap'' Cooper pairs, and that these features seen in the data don't arise from some interplay between the different width leads in our devices.  Our geometry could even be insufficient to constrain the Cooper pairs.  In the smallest samples, the ratio between the width of the thin leads, and the size of the square is only $\sim$30\% meaning that the squares cannot fully confine the Cooper pairs, and they will leak into the leads.  This means our observations could be due to a consequence of our experimental design rather than effects originating from confining Cooper pairs.  In addition, our experimental design complicates the analysis because the two thin leads are expected to possess enhanced condensation energy, and thus the critical field as well\cite{deGennesLoop}.  Near T$_c$ the coherence length diverges, and thus the sample also becomes coupled to the wider leads which have a reduced H$_c$ decreasing the observed H$_c$.  Noting that samples C, D and E were not four-point resistivity measurements, therefore the phase diagram constructed at a particular resistance value may have included the resistance of the sample as well as the resistance of the leads.  With all this in mind, we propose to explain the nature of the observed features using proximity effect between the different width lead sections having different H$_c$ along the device, see inset of Fig. 4.  However, the absence of the ``quantized'' behavior in sample C which differs from sample D only in $\xi$(0) indicates that our observed features are not explained by a simple interplay between the leads and square.   More experiments, with better controlled geometries and possibly tunneling electrodes are needed to resolve these issues.  

We would like to acknowledge useful discussions with J. Sauls, X. Hu, Y. A. Ying, and C. P. Puls. This work is supported by NSF under Grant DMR 0908700 and Penn State MRI Nanofabrication Lab under NSF Cooperative Agreement 0335765, NNIN with Cornell University.
\bibliography{Bib.bib}

\begin{thebibliography}{18}
\expandafter\ifx\csname natexlab\endcsname\relax\def\natexlab#1{#1}\fi
\expandafter\ifx\csname bibnamefont\endcsname\relax
  \def\bibnamefont#1{#1}\fi
\expandafter\ifx\csname bibfnamefont\endcsname\relax
  \def\bibfnamefont#1{#1}\fi
\expandafter\ifx\csname citenamefont\endcsname\relax
  \def\citenamefont#1{#1}\fi
\expandafter\ifx\csname url\endcsname\relax
  \def\url#1{\texttt{#1}}\fi
\expandafter\ifx\csname urlprefix\endcsname\relax\def\urlprefix{URL }\fi
\providecommand{\bibinfo}[2]{#2}
\providecommand{\eprint}[2][]{\url{#2}}

\bibitem[{\citenamefont{Anderson}(1959)}]{Anderson59}
\bibinfo{author}{\bibfnamefont{P.~W.} \bibnamefont{Anderson}},
  \bibinfo{journal}{J. Phys. Chem. Solids} \textbf{\bibinfo{volume}{11}},
  \bibinfo{pages}{28} (\bibinfo{year}{1959}).

\bibitem[{\citenamefont{Tinkham}(2000)}]{TinkhamReview}
\bibinfo{author}{\bibfnamefont{M.}~\bibnamefont{Tinkham}},
  \bibinfo{journal}{Journal of Superconductivity}
  \textbf{\bibinfo{volume}{13}}, \bibinfo{pages}{801} (\bibinfo{year}{2000}).

\bibitem[{\citenamefont{Black et~al.}(1996)\citenamefont{Black, Ralph, and
  Tinkham}}]{BlackTinkhamNmAl}
\bibinfo{author}{\bibfnamefont{C.~T.} \bibnamefont{Black}},
  \bibinfo{author}{\bibfnamefont{D.~C.} \bibnamefont{Ralph}}, \bibnamefont{and}
  \bibinfo{author}{\bibfnamefont{M.}~\bibnamefont{Tinkham}},
  \bibinfo{journal}{Phys.\ Rev.\ Lett.} \textbf{\bibinfo{volume}{76}},
  \bibinfo{pages}{688} (\bibinfo{year}{1996}).

\bibitem[{\citenamefont{Buisson et~al.}(1990)\citenamefont{Buisson, Gandit,
  Rammal, Wang, and Pannetier}}]{BuissonPhysLetA}
\bibinfo{author}{\bibfnamefont{O.}~\bibnamefont{Buisson}},
  \bibinfo{author}{\bibfnamefont{P.}~\bibnamefont{Gandit}},
  \bibinfo{author}{\bibfnamefont{R.}~\bibnamefont{Rammal}},
  \bibinfo{author}{\bibfnamefont{Y.~Y.} \bibnamefont{Wang}}, \bibnamefont{and}
  \bibinfo{author}{\bibfnamefont{B.}~\bibnamefont{Pannetier}},
  \bibinfo{journal}{Physics Letters A} \textbf{\bibinfo{volume}{150}},
  \bibinfo{pages}{36} (\bibinfo{year}{1990}).

\bibitem[{\citenamefont{Moshcalkov et~al.}(1995)\citenamefont{Moshcalkov,
  Gielen, Strunk, Jonckheere, Qiu, Haesendonck, and
  Bruynseraede}}]{MoshcalkovNature95}
\bibinfo{author}{\bibfnamefont{V.~V.} \bibnamefont{Moshcalkov}},
  \bibinfo{author}{\bibfnamefont{L.}~\bibnamefont{Gielen}},
  \bibinfo{author}{\bibfnamefont{C.}~\bibnamefont{Strunk}},
  \bibinfo{author}{\bibfnamefont{R.}~\bibnamefont{Jonckheere}},
  \bibinfo{author}{\bibfnamefont{X.}~\bibnamefont{Qiu}},
  \bibinfo{author}{\bibfnamefont{C.~V.} \bibnamefont{Haesendonck}},
  \bibnamefont{and}
  \bibinfo{author}{\bibfnamefont{Y.}~\bibnamefont{Bruynseraede}},
  \bibinfo{journal}{Nature} \textbf{\bibinfo{volume}{373}},
  \bibinfo{pages}{319} (\bibinfo{year}{1995}).

\bibitem[{\citenamefont{Geim et~al.}(1997)\citenamefont{Geim, Grigorieva,
  Dubonos, Lok, Maan, Filippov, and Peeters}}]{GeimNature97}
\bibinfo{author}{\bibfnamefont{A.~K.} \bibnamefont{Geim}},
  \bibinfo{author}{\bibfnamefont{I.~V.} \bibnamefont{Grigorieva}},
  \bibinfo{author}{\bibfnamefont{S.~V.} \bibnamefont{Dubonos}},
  \bibinfo{author}{\bibfnamefont{J.~G.~S.} \bibnamefont{Lok}},
  \bibinfo{author}{\bibfnamefont{J.~C.} \bibnamefont{Maan}},
  \bibinfo{author}{\bibfnamefont{A.~E.} \bibnamefont{Filippov}},
  \bibnamefont{and} \bibinfo{author}{\bibfnamefont{F.~M.}
  \bibnamefont{Peeters}}, \bibinfo{journal}{Nature}
  \textbf{\bibinfo{volume}{390}}, \bibinfo{pages}{1997} (\bibinfo{year}{1997}).

\bibitem[{\citenamefont{Chibotaru et~al.}(2000)\citenamefont{Chibotaru,
  Ceulemans, Bruyndoncx, and Moshcalkov}}]{ChibotaruNatureAntivortex}
\bibinfo{author}{\bibfnamefont{L.~F.} \bibnamefont{Chibotaru}},
  \bibinfo{author}{\bibfnamefont{A.}~\bibnamefont{Ceulemans}},
  \bibinfo{author}{\bibfnamefont{V.}~\bibnamefont{Bruyndoncx}},
  \bibnamefont{and} \bibinfo{author}{\bibfnamefont{V.~V.}
  \bibnamefont{Moshcalkov}}, \bibinfo{journal}{Nature}
  \textbf{\bibinfo{volume}{408}}, \bibinfo{pages}{833} (\bibinfo{year}{2000}).

\bibitem[{\citenamefont{Nishio et~al.}(2008)\citenamefont{Nishio, An, Nomura,
  Miyachi, Eguchi, Sakata, Lin, Hayashi, Nakai, Machida et~al.}}]{NishioSTMPb}
\bibinfo{author}{\bibfnamefont{T.}~\bibnamefont{Nishio}},
  \bibinfo{author}{\bibfnamefont{T.}~\bibnamefont{An}},
  \bibinfo{author}{\bibfnamefont{A.}~\bibnamefont{Nomura}},
  \bibinfo{author}{\bibfnamefont{K.}~\bibnamefont{Miyachi}},
  \bibinfo{author}{\bibfnamefont{T.}~\bibnamefont{Eguchi}},
  \bibinfo{author}{\bibfnamefont{H.}~\bibnamefont{Sakata}},
  \bibinfo{author}{\bibfnamefont{S.}~\bibnamefont{Lin}},
  \bibinfo{author}{\bibfnamefont{N.}~\bibnamefont{Hayashi}},
  \bibinfo{author}{\bibfnamefont{N.}~\bibnamefont{Nakai}},
  \bibinfo{author}{\bibfnamefont{M.}~\bibnamefont{Machida}},
  \bibnamefont{et~al.}, \bibinfo{journal}{Phys.\ Rev.\ Lett.}
  \textbf{\bibinfo{volume}{101}}, \bibinfo{pages}{167001}
  (\bibinfo{year}{2008}).

\bibitem[{\citenamefont{Wang et~al.}(2009)\citenamefont{Wang, Zhang, Loy,
  Chiang, and Xiao}}]{WangPseudogapPbSTM}
\bibinfo{author}{\bibfnamefont{K.}~\bibnamefont{Wang}},
  \bibinfo{author}{\bibfnamefont{X.}~\bibnamefont{Zhang}},
  \bibinfo{author}{\bibfnamefont{M.~M.~T.} \bibnamefont{Loy}},
  \bibinfo{author}{\bibfnamefont{T.~C.} \bibnamefont{Chiang}},
  \bibnamefont{and} \bibinfo{author}{\bibfnamefont{X.}~\bibnamefont{Xiao}},
  \bibinfo{journal}{Phys.\ Rev.\ Lett.} \textbf{\bibinfo{volume}{102}},
  \bibinfo{pages}{076801} (\bibinfo{year}{2009}).

\bibitem[{\citenamefont{Bose et~al.}(2010)\citenamefont{Bose, Garcia-Garcia,
  Ugeda, Urbina, Michaelis, Brihuega, and Kern}}]{BoseShellEffectsSnSTM}
\bibinfo{author}{\bibfnamefont{S.}~\bibnamefont{Bose}},
  \bibinfo{author}{\bibfnamefont{A.~M.} \bibnamefont{Garcia-Garcia}},
  \bibinfo{author}{\bibfnamefont{M.~M.} \bibnamefont{Ugeda}},
  \bibinfo{author}{\bibfnamefont{J.~D.} \bibnamefont{Urbina}},
  \bibinfo{author}{\bibfnamefont{C.~H.} \bibnamefont{Michaelis}},
  \bibinfo{author}{\bibfnamefont{I.}~\bibnamefont{Brihuega}}, \bibnamefont{and}
  \bibinfo{author}{\bibfnamefont{K.}~\bibnamefont{Kern}},
  \bibinfo{journal}{Nature Materials} \textbf{\bibinfo{volume}{9}},
  \bibinfo{pages}{550} (\bibinfo{year}{2010}).

\bibitem[{\citenamefont{Little and Parks}(1962)}]{LittleParks}
\bibinfo{author}{\bibfnamefont{W.~A.} \bibnamefont{Little}} \bibnamefont{and}
  \bibinfo{author}{\bibfnamefont{R.~D.} \bibnamefont{Parks}},
  \bibinfo{journal}{Phys.\ Rev.\ Lett.} \textbf{\bibinfo{volume}{9}},
  \bibinfo{pages}{9} (\bibinfo{year}{1962}).

\bibitem[{\citenamefont{de~Gennes}(1981)}]{deGennesLoop}
\bibinfo{author}{\bibfnamefont{P.~G.} \bibnamefont{de~Gennes}},
  \bibinfo{journal}{C. R. Acad. Sci. Ser. II} \textbf{\bibinfo{volume}{292}},
  \bibinfo{pages}{279} (\bibinfo{year}{1981}).

\bibitem[{\citenamefont{Liu et~al.}(2001)\citenamefont{Liu, Zadorozhny,
  Rosario, Rock, Carrigan, and Wang}}]{LiuScience01}
\bibinfo{author}{\bibfnamefont{Y.}~\bibnamefont{Liu}},
  \bibinfo{author}{\bibfnamefont{Y.}~\bibnamefont{Zadorozhny}},
  \bibinfo{author}{\bibfnamefont{M.~M.} \bibnamefont{Rosario}},
  \bibinfo{author}{\bibfnamefont{B.~Y.} \bibnamefont{Rock}},
  \bibinfo{author}{\bibfnamefont{P.~T.} \bibnamefont{Carrigan}},
  \bibnamefont{and} \bibinfo{author}{\bibfnamefont{H.}~\bibnamefont{Wang}},
  \bibinfo{journal}{Science} \textbf{\bibinfo{volume}{294}},
  \bibinfo{pages}{2332} (\bibinfo{year}{2001}).

\bibitem[{\citenamefont{Schweigert and Peeters}(1998)}]{SchweigertPeeters98}
\bibinfo{author}{\bibfnamefont{V.~A.} \bibnamefont{Schweigert}}
  \bibnamefont{and} \bibinfo{author}{\bibfnamefont{F.~M.}
  \bibnamefont{Peeters}}, \bibinfo{journal}{Phys. Rev. B}
  \textbf{\bibinfo{volume}{57}}, \bibinfo{pages}{13817} (\bibinfo{year}{1998}).

\bibitem[{\citenamefont{de~Gennes}(1999)}]{DeGennes}
\bibinfo{author}{\bibfnamefont{P.}~\bibnamefont{de~Gennes}},
  \emph{\bibinfo{title}{Superconductivity of Metals and Alloys}}
  (\bibinfo{publisher}{Westview Press}, \bibinfo{year}{1999}).

\bibitem[{\citenamefont{Mertelj and Kabanov}(2003)}]{MerteljVortexPhaseDiagram}
\bibinfo{author}{\bibfnamefont{T.}~\bibnamefont{Mertelj}} \bibnamefont{and}
  \bibinfo{author}{\bibfnamefont{V.~V.} \bibnamefont{Kabanov}},
  \bibinfo{journal}{Phys. Rev. B} \textbf{\bibinfo{volume}{67}},
  \bibinfo{pages}{134527} (\bibinfo{year}{2003}).

\bibitem[{\citenamefont{Benoist and Zwerger}(1997)}]{BenoistZwerger96}
\bibinfo{author}{\bibfnamefont{R.}~\bibnamefont{Benoist}} \bibnamefont{and}
  \bibinfo{author}{\bibfnamefont{W.}~\bibnamefont{Zwerger}},
  \bibinfo{journal}{Z. Phys. B} \textbf{\bibinfo{volume}{103}},
  \bibinfo{pages}{377} (\bibinfo{year}{1997}).

\bibitem[{\citenamefont{Fink and Gr\"{u}nfeld}(1985)}]{FinkSideBranch}
\bibinfo{author}{\bibfnamefont{H.~J.} \bibnamefont{Fink}} \bibnamefont{and}
  \bibinfo{author}{\bibfnamefont{V.}~\bibnamefont{Gr\"{u}nfeld}},
  \bibinfo{journal}{Phys.\ Rev.\ B} \textbf{\bibinfo{volume}{31}},
  \bibinfo{pages}{600} (\bibinfo{year}{1985}).

\end{thebibliography}

\end{document}